\documentclass{sig-alternate-per}

\usepackage[noend]{algpseudocode}
\usepackage{algorithm}
\usepackage{braket}
\usepackage{xcolor}

\begin{document}

\title{{Noise estimation in an entanglement distillation protocol}}

\numberofauthors{3}
\author{
\alignauthor
Ananda G. Maity$^\text{*}$
\\
       \affaddr{Networked Quantum Devices Unit, Okinawa Institute of Science and Technology Graduate University,}\\
       \affaddr{Onna-son, Okinawa 904-0495, Japan}\\
\alignauthor
Joshua C. A. Casapao
\\
      \affaddr{Networked Quantum Devices Unit, Okinawa Institute of Science and Technology Graduate University,}\\
       \affaddr{ Onna-son, Okinawa 904-0495, Japan}\\
\alignauthor  Naphan Benchasattabuse
\\
       \affaddr{Keio University Shonan Fujisawa Campus,}\\
       \affaddr{5322 Endo, Fujisawa, Kanagawa
252-0882, Japan}\\
\and 
\alignauthor  Michal Hajdu\v{s}ek\\
       \affaddr{Keio University Shonan Fujisawa Campus,}\\
       \affaddr{5322 Endo, Fujisawa, Kanagawa
252-0882, Japan}\\
\alignauthor Rodney Van Meter\\
        \affaddr{Keio University Shonan Fujisawa Campus,}\\
       \affaddr{5322 Endo, Fujisawa, Kanagawa
252-0882, Japan}\\
\alignauthor David Elkouss\\
       \affaddr{Networked Quantum Devices Unit, Okinawa Institute of Science and Technology Graduate University,}\\
       \affaddr{ Onna-son, Okinawa 904-0495, Japan}\\
}

\maketitle
\begin{abstract}
Estimating noise processes is an essential step for practical quantum information processing. Standard estimation tools require consuming valuable quantum resources. Here we ask the question of whether the noise affecting entangled states can be learned solely from the measurement statistics obtained during a distillation protocol. As a first step, we consider states of the Werner form and find that the Werner parameter can be estimated efficiently from the measurement statistics of an idealized distillation protocol. Our proposed estimation method can find application in scenarios where distillation is an unavoidable step.
\end{abstract}


\section{Introduction}
\vspace{0.4cm}
The quantum Internet will provide an avenue for creating quantum information shared among distant nodes of a network, allowing distributed information processing and communication tasks which are beyond the capabilities of classical networks~\cite{Wehner18,VanMeter22,rfc9340}. Some notable examples are: quantum key distribution, higher-precision clock synchronization, and distributed quantum computing.

However, realisation of the above tasks requires the distribution of high quality entanglement between the distant nodes. Long-distance remote entanglement distribution remains a challenge as entanglement attenuates exponentially with distance and cannot be amplified because of the no-cloning theorem; quantum repeaters \cite{azuma22} and entanglement distillation protocols \cite{Bennett96(1)} can be used to overcome the effects of noise and loss. In particular, an entanglement distillation protocol transforms a number of low-quality entangled states into a smaller number of entangled states with higher quality.

Thus, designing methods for estimating the noise processes efficiently is of foremost importance. Approaches include tomography, randomized benchmarking, self-testing, quantum gate set tomography and so on~\cite{Eisert2020}. However, most of these estimation methods fully measure the quantum resources, making them useless for further information processing tasks. 

In this work, we ask whether the noise parameters can be efficiently characterized from the measurement statistics obtained during a distillation protocol. This is particularly relevant for those information processing tasks where distillation is an unavoidable step. For example in a quantum network, for establishing high quality entanglement, one may need to perform distillation. Moreover, eliminating noise estimation as a separate task might also simplify the quantum network management.

In the following, we describe a procedure for estimating the noise parameter from the measurement statistics of a distillation protocol. We investigate the number of samples required for estimating the state parameters of the Werner state in a usual tomography method and in a distillation protocol. 
We find that the distillation-based estimation protocol can be more efficient than state tomography for some parameter regimes. 
We also propose an algorithm for experimental realisation of the distillation based estimation protocol.

\section{Preliminaries}
\vspace{0.4cm}
\textbf{Distillation protocol:} We consider the distillation protocol proposed by Bennett {\it et. al} \cite{Bennett96(1)}, where two distant parties (Alice and Bob) cooperate to improve the fidelity of shared quantum states. The protocol proceeds as follows:

1) Alice and Bob start with two copies of a noisy entangled state. Here, we assume that the state is a Werner state and characterized by  the Werner parameter $w$:
\begin{align}\label{Werner_state}
    \rho_{w}=& (1-w)\ket{\Phi^+}\bra{\Phi^+} + \frac{w}{4} \mathbb{I}
\end{align}
The fidelity of such a Werner state as a function of $w$ with $\ket{\Phi^+}$ is $F= 1-\frac{3w}{4}$.

2) Each applies local XOR operations on the local parts of their two copies, with one copy as the control and another one as the target.

3) Each of them measures the target qubit in the $Z$-basis and communicates the results classically.

4) If they obtain correlated outcomes i.e, $00,11$ they keep the unmeasured copy, otherwise they discard the state. 

After doing this protocol, the fidelity of the new state becomes $ F' = \frac{F^2 + \frac{1}{9}(1-F)^2}{F^2 + \frac{2}{3}F(1-F) + \frac{5}{9}(1-F)^2}$ where $F' > F$ if $F >\frac{1}{2}$ \cite{Bennett96(1)}. One may in general repeat the above protocol recursively and obtain a perfect $\ket{\Phi^+}$ asymptotically.

\textbf{Statistical tools used for the parameter estimation:} An important question in probability theory is: given a random variable $X$ and its expectation $\mathbb{E} (X)$, how likely is $X$ to be close to $\mathbb{E} (X)$ i.e, $\text{Pr}(|X-\mathbb{E} (X)| \geq t)$ for some $t\geq 0$? For this purpose, here we use Hoeffding's inequality. Suppose $X_1,...,X_n$ are independent random variables and $a \leq X_i \leq b$, then for any $t > 0$, 
\begin{equation}
    \text{Pr}(\vert\frac{1}{n}\sum_i(X_i-\mathbb{E}(X_i) \vert \geq t) \leq 2 \exp{\left(-\frac{2nt^2}{(b -a)^2}\right)}.
\end{equation}

\section{Werner parameter estimation}
\vspace{0.4cm}
Given an unknown state of the Werner form, here we sketch how to estimate the Werner parameter from the measurement outcomes of a distillation protocol. To evaluate the potential interest of this approach, we compare the precision of the estimate with the precision achieved with state tomography.

\textbf{Werner parameter estimation in a distillation protocol:} Assume that Alice and Bob hold two copies of a two-qubit Werner state and perform the distillation protocol described in the previous section. The probability of observing the outcome $00$, $p_{00}$, can be computed as $p_{00} = \frac{1}{4}(2-2w+ w^2)$. Or alternatively;
\begin{equation} 
\label{eq:wfromp00}
w = 1- \sqrt{4p_{00}-1}.
\end{equation}
The value of $p_{00}$ also corresponds to the expected value of a random variable $P_{00}$ that assigns value 1 to the measurement outcomes $00$ and value 0 to any other outcome i.e, $p_{00} = \mathbb{E} (P_{00})$.

Now in an experiment, one may estimate the probability of observing the outcomes 00. We denote this estimate by $\hat{p}_{00}$. Then from Hoeffding's inequality, we have
\begin{equation}
    \text{Pr}(|\hat{p}_{00} - {p}_{00}|\geq \epsilon) \leq 2\exp{(-{2}n\epsilon^2)}.
\end{equation}

\begin{figure}[!h] 
\includegraphics[width=0.5\textwidth]{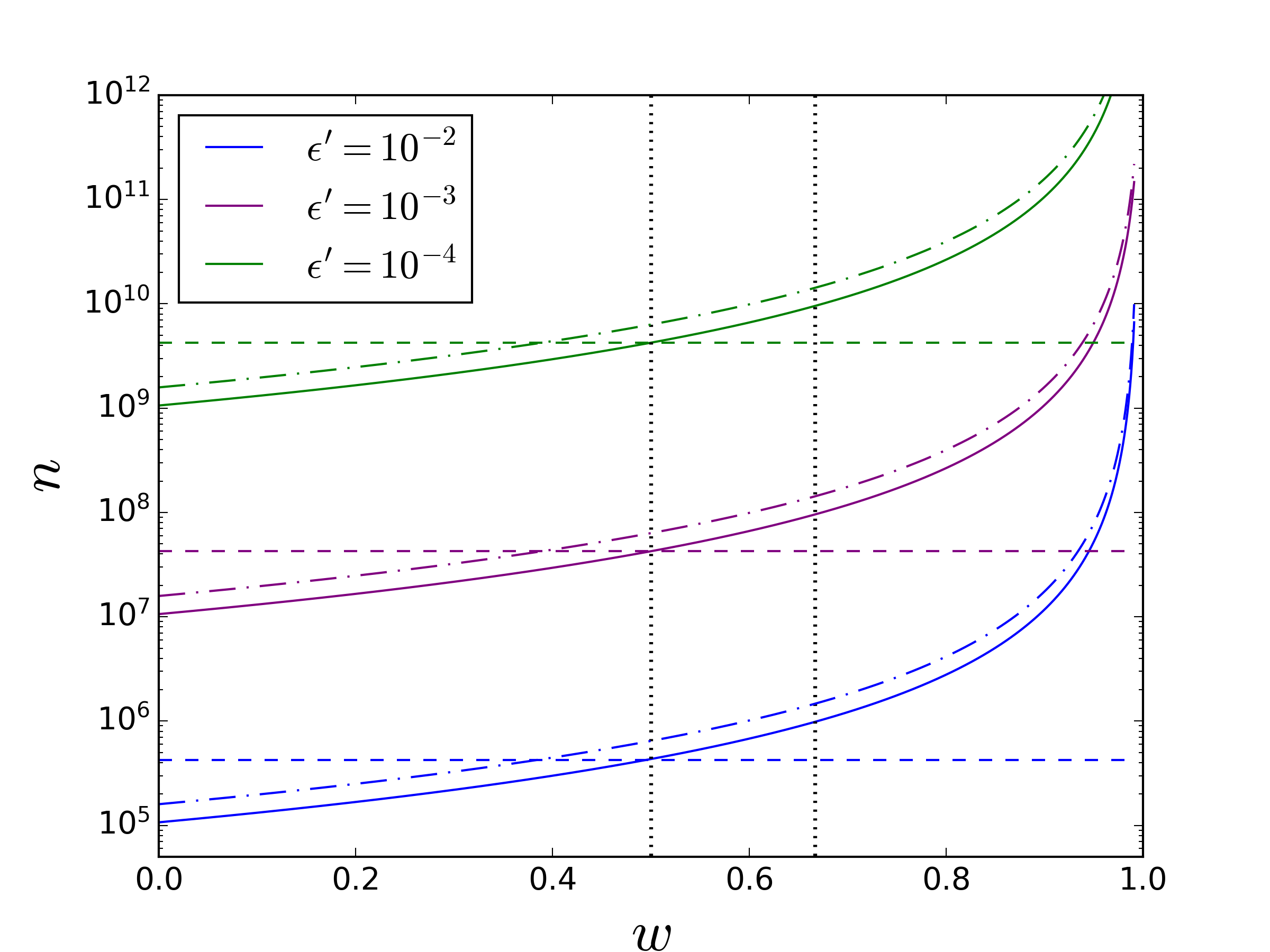}
\caption{Minimum number of samples $n$ required for estimating the Werner parameter $w$ with failure probability $\mathrm{Pr}(|\hat{w} - w|\geq \epsilon') = 0.01$. The solid (dashed) curves correspond to the estimation via a noise-free distillation protocol (via tomography). The dash-dotted curves correspond to the distillation protocol in the presence of depolarizing noise, where $S = \exp(-1/5)$. {For the range $w \leq 1/2$}, we find that the number of samples required for a successful distillation-based estimation can be significantly less than that for tomography. }\label{fig1}
\centering 
\end{figure}

From $\hat{p}_{00}$, it is possible to obtain an estimate $\hat w$ of the Werner parameter with Equation \ref{eq:wfromp00}. Let us now sketch how to bound the precision of the estimated $\hat{w}$. If we desire to have $|\hat{w}- w| \geq \epsilon'$ then from Equation \ref{eq:wfromp00}, we obtain $|\sqrt{4p_{00}-1} - \sqrt{4\hat{p}_{00}-1}| \geq \epsilon'$. After explicit calculation above condition implies either $\hat{p}_{00}-p_{00} \geq \frac{1}{4} (-\epsilon'^{2} + 2\epsilon'(1-w))$ or $-(\hat{p}_{00}-p_{00}) \geq \frac{1}{4} (\epsilon'^{2} + 2\epsilon'(1-w))$. Taking the loose bound, we can conclude that when $|\hat{w}- w| \geq \epsilon'$, then $|\hat{p}_{00} - p_{00}| \geq \frac{1}{4} (-\epsilon'^{2} + 2\epsilon'(1-w))$. Consequently, $\text{Pr}(|\hat{w}- w| \geq \epsilon')$ implies $\text{Pr}(|\hat{p}_{00} - p_{00}|\geq  \frac{1}{4} (-\epsilon'^{2} + 2\epsilon'(1-w)))$. Using Hoeffding's inequality leads to the bound
\begin{equation} \label{estimation_werner}
    \text{Pr}(|\hat{w}- w| \geq \epsilon') \leq 2\exp{(-{\frac{1}{8}  } n (-\epsilon'^{2} + 2\epsilon'(1-w)){^2})}.
\end{equation}
We investigate the scaling of the number of samples with respect to the Werner parameter $w$ when $\text{Pr}(|\hat{w}- w| \geq \epsilon') = 0.01$ in Fig.~\ref{fig1}.

\textbf{Werner parameter estimation in tomography:}
A Werner state can be written in the locally decomposable form $\rho_{w}= \frac{1}{4}[\mathbb{I}\otimes \mathbb{I} + (1-w)(X \otimes X - Y \otimes Y + Z \otimes Z)]$.

In order to do tomography of a bipartite entangled state, Alice and Bob can measure locally $\lbrace \mathbb{I},X,Y,Z \rbrace$ and depending on the joint probability of $X \otimes X$, $Y \otimes Y$, $Z \otimes Z$ they can estimate the Werner parameter. One may calculate the probability of getting $00$ (or $11$) outcome, (when both of them measure in the $Z$ basis) $ p_{00} = p_{11}= \frac{2-w}{4}$. Now if $|\hat{w}- w| \geq \epsilon'$, then one may note that $|\hat{p}_{00} - p_{00}|\geq \frac{\epsilon'}{4}$. Using Hoeffding's inequality, we get the bound
\begin{equation}\label{tomography_Hoff}
    \text{Pr}(|\hat{w}- w| \geq \epsilon') \leq 2\exp{(-{\frac{1}{8}  }n\epsilon'^{2})}.
\end{equation}

\textbf{Estimation of Werner parameter in the presence of depolarizing noise:} As a first step towards the application of these ideas in a realistic setting, we consider that idling qubits undergo depolarizing noise with the depolarizing operation given by 
\begin{equation}
    \Lambda (\rho) = (1-x)\rho + \frac{x}{4} \mathbb{I}.
\end{equation}
Here $x= 1- e^{-\frac{t}{T}}$ is the depolarizing parameter. The distillation protocol under consideration requires two copies of the states. We model the entanglement generation as a sequential process: after the first copy is prepared, it undergoes depolarizing noise. However, the second copy can be used for distillation as soon as it is generated and hence may be considered to be noise free. We follow the same procedure as discussed earlier and estimate the Werner parameter. Explicit calculation reveals that if $|\hat{w}- w|  \geq \epsilon'$, then $|\hat{p}_{00} - p_{00}| \geq  \frac{S}{4} (-\epsilon'^{2} + 2\epsilon'(1-w))$ with $S= \frac{1}{n}\sum_{i=1}^{n} (1-x_{i})$. Hoeffding's inequality leads to the bound
\begin{equation}
    \text{Pr}(|\hat{w}- w|\geq \epsilon') \leq 2\exp{(- {\frac{1}{8}  }n S^2( 2\epsilon'(1-w)-\epsilon'^{2} ){^2})}
\end{equation}
Note that for tomography the states can be used as soon as they are generated and hence can be considered as noise free. Thus the bound on the number of samples is independent of the depolarizing noise parameter.

\section{Algorithmic estimation}
\vspace{0.4cm}
Here we provide an algorithmic procedure that can be used in an idealized experiment for estimating the Werner parameter. The overall process is described in Algorithm 1.

The input for this procedure is the number of samples $N$ and the target estimate precision $\epsilon_w$.

For each sample, Alice and Bob prepare two 
copies of the entangled state. Then, they  apply a local XOR operation on their respective local copies and finally measure the second copy in the $Z$ basis. 

After repeating the experiment $N$ times, they collect the measurement statistics and estimate the probability of obtaining outcome $00$ as $\hat{p}_{00}$. From $\hat{p}_{00}$, Alice and Bob can estimate $\hat{w}$ as $\hat{w}=1-\sqrt{4\hat{p}_{00}-1}$.

The next step is to translate the target precision $\epsilon_w$ to a precision on $\hat p_{00}$. The goal of Alice and Bob is to estimate the probability that $\hat{w}$ is in the range $[w_m,w_p]=[\hat{w} - \epsilon_{w}, \hat{w} + \epsilon_{w}]$. Using the relation between the Werner parameter and $p_{00}$, this probability is equivalent to the probability that $\hat p_{00}$ is in the range $[p_{00m},p_{00p}]=[\frac{1}{4} (2- 2w_{m}+w_{m}^{2}),\frac{1}{4} (2- 2w_{p}+w_{p}^{2})]$. We can finally let pessimistically $\epsilon = \text{Max} [|\hat{p}_{00}- p_{00p}|, |\hat{p}_{00}- p_{00m}|]$ and from Hoeffding's inequality we conclude that $\text{Pr}(|\hat{w}- w|\geq \epsilon_{w})\leq 2 \exp (-{2}N\epsilon^{2})$.

\begin{figure}
\begin{minipage}{\linewidth}
\begin{algorithm}[H]
{\small
\begin{algorithmic}[1]
\caption{\small Experiment for estimation of Werner parameter.}
\label{alg:U}
\Require 
\Statex number of samples $N$
\Statex estimation precision $\epsilon_w$
\Ensure 
\Statex failure probability $\delta$
\Statex estimated Werner parameter $\hat{w}$
\State $n_\text{count} \leftarrow 0$
\For{$n=1$ to $N$}
\State Prepare two copies of the unknown state $\rho_{w, c}$ and $\rho_{w, t}$
\State Alice and Bob locally perform XOR-operation
\State $(Z_A, Z_B) \leftarrow M_Z^{\otimes 2} \rho_{w, t}$ \Comment{Each measures their half of $\rho_{w, t}$ in the $Z$-basis}
\If{$Z_A = +1$ and $Z_B = +1$}
\State $n_\text{count} \leftarrow n_\text{count}+1$
\EndIf
\EndFor
\State $\hat{p}_{00} \leftarrow n_\text{count}/N$
\State $\hat{w} \leftarrow 1-\sqrt{4\hat{p}_{00}-1}$
\Statex (Calculate the extreme values around $\hat{w}$)
\State $w_p \leftarrow \hat{w} + \epsilon_{w}$
\State $w_m \leftarrow \hat{w}  - \epsilon_{w}$
\Statex (Calculate the extreme values around $p_{00}$)
\State $p_{00p} \leftarrow \frac{1}{4} (2- 2w_{p}+w_{p}^{2})$
\State $p_{00m} \leftarrow \frac{1}{4} (2- 2w_{m}+w_{m}^{2})$
\Statex (Calculate precision around $p_{00}$)
\State $\epsilon \leftarrow \text{max} (|\hat{p}_{00}- p_{00p}|, |\hat{p}_{00}- p_{00m}|)$
\State $\delta \leftarrow 2\exp (-{2}N\epsilon^{2})$. 
\State Output $\delta$, $\hat{w}$
\end{algorithmic}
}
\end{algorithm}
\end{minipage}
\end{figure}

\section{Conclusion}
\vspace{0.4cm}
Noise estimation is a key requirement for quantum technologies. In the case of quantum networks, precise entanglement characterization allows us to optimize task implementation or to maximize the end-to-end fidelity with entanglement routing protocols. Existing methods consume valuable entanglement for the purpose of estimation. In this work, we have shown that in an idealized scenario, the measurement outcomes of distillation protocols allow us to characterize efficiently an entangled state. In particular, whenever distillation is a necessary step, this method works without consuming any quantum resources. However, the models for noise and processes considered here are ideal. Further work is necessary to establish the practical relevance of this approach.
\section{Acknowledgements}
\vspace{0.4cm}
This work was supported  by the JST Moonshot R\&D program under Grants JPMJMS226C.
\bibliographystyle{acm}
\vspace{0.4cm}
\bibliography{Biblio}

\begin{thebibliography}{1}

\bibitem{azuma22}
{\sc Azuma, K., Economou, S.~E., Elkouss, D., Hilaire, P., Jiang, L., Lo,
  H.-K., and Tzitrin, I.}
\newblock Quantum repeaters: From quantum networks to the quantum internet.
\newblock {\em Rev. Mod. Phys. 95\/} (Dec 2023), 045006.

\bibitem{Bennett96(1)}
{\sc Bennett, C.~H., Brassard, G., Popescu, S., Schumacher, B., Smolin, J.~A.,
  and Wootters, W.~K.}
\newblock Purification of noisy entanglement and faithful teleportation via
  noisy channels.
\newblock {\em Phys. Rev. Lett. 76\/} (Jan 1996), 722--725.

\bibitem{Eisert2020}
{\sc Eisert, J., Hangleiter, D., Walk, N., Roth, I., Markham, D., Parekh, R.,
  Chabaud, U., and Kashefi, E.}
\newblock Quantum certification and benchmarking.
\newblock {\em Nature Reviews Physics 2}, 7 (June 2020), 382--390.

\bibitem{rfc9340}
{\sc Kozlowski, W., Wehner, S., Meter, R.~V., Rijsman, B., Cacciapuoti, A.~S.,
  Caleffi, M., and Nagayama, S.}
\newblock {Architectural Principles for a Quantum Internet}.
\newblock RFC 9340, Mar. 2023.

\bibitem{VanMeter22}
{\sc Van~Meter, R., Satoh, R., Benchasattabuse, N., Teramoto, K., Matsuo, T.,
  Hajdušek, M., Satoh, T., Nagayama, S., and Suzuki, S.}
\newblock A quantum internet architecture.
\newblock In {\em 2022 IEEE International Conference on Quantum Computing and
  Engineering (QCE)\/} (2022), pp.~341--352.

\bibitem{Wehner18}
{\sc Wehner, S., Elkouss, D., and Hanson, R.}
\newblock Quantum internet: A vision for the road ahead.
\newblock {\em Science 362}, 6412 (2018), eaam9288.

\end{thebibliography}
\end{document}